# Effect of the selective localization of carbon nanotubes and phase domain in immiscible blends on tunable microwave dielectric properties

Liping Zhou[1], Yu Tian[1], Peng Xu[1], Huijie Wei[1], Yuhan Li[2], Hua-Xin Peng[1], Faxiang Qin[1*]

[1]*Institute for Composites Science Innovation (InCSI), School of Materials Science and Engineering, Zhejiang University, 38 Zheda Road, Hangzhou, 310027, PR. China*

[2]*Department of Materials, Imperial College London, Exhibition Road, London, SW7 2AZ, United Kingdom*

**Abstract**

In recent years, the immiscible polymer blend system has attracted much attention as the matrix of nanocomposites. Herein, from the perspective of dynamics, the control of the carbon nanotubes (CNTs) migration aided with the interface of polystyrene (PS) and poly(methyl methacrylate) (PMMA) blends was achieved through a facile melt mixing method. Thus, we revealed a comprehensive relationship between several typical CNTs migrating scenarios and the microwave dielectric properties of their nanocomposites. Based on the unique morphologies and phase domain structures of the immiscible matrix, we further investigated the multiple microwave dielectric relaxation processes and shed new light on the relation between relaxation peak position and the phase domain size distribution. Moreover, by integrating the CNTs interface localization control with the matrix co-continuous structure construction, we found that the interface promotes double percolation effect to achieve conductive percolation at low CNTs loading (~1.06 vol%). Overall, the present study provides a unique nanocomposite material design symphonizing both functional fillers dispersion and location as well as the matrix architecture optimization for microwave applications.

[*] Corresponding author: Tel: 187 6718 9652. E-mail: faxiangqin@zju.edu.cn (F. Qin).

## 1. Introduction

With the emergence and continuous development of nanotechnology, nanocomposites have exhibited many excellent properties[1]. Among others, dielectric nanocomposites have garnered much research interest. The dielectric properties of nanocomposites play a key role in the development of their applications as dielectric materials (energy storage devices, wave absorbing elements, transducers, *etc*.)[2]. Traditionally, increasing the content of the nanofillers or using multi-fillers can generally improve the performance of the nanocomposites[3]. However, under such circumstance, the complexity of material design and structural incompleteness also increase correspondingly, and the lightweight feature and excellent processability of nanocomposites would be compromised[4]. Consequently, we proposed the concept of "plainified composites" in our previous research which refers to optimization of the comprehensive performance without a change in the component of composites through the design and control of the multi-level structure of the composites (interface, dispersion and distribution of functional phases, and macrostructure design, etc.) without altering the composition of the material system. Our previous work, for instance, demonstrated that, through engineering peculiar interfaces, we were able to realize desirable material performance[5]. We introduced a vertical interface composed of two phases with different polarization abilities across their interface in carbon nanotube/silicone elastomer nanocomposites and this proposed method induce a distinguishable dielectric response, which shows great potential in enhancing the high frequency dielectric response of nanocomposites and expanding the tunability of electromagnetic properties. In this paper, following the concept of "plainified composites", we introduce an immiscible polymer blend system to further increase the flexibility of dielectric functional nanocomposites. We regulate the electric properties of nanocomposites through a design of the two-phase interface of the immiscible blend and the control of the dispersion and distribution of functional phase, without the need for excessive addition and complicated chemical processing.

Polymers are typically immiscible and usually acquire a small-scale arrangement of the phases when mixed together because of the unfavorably low mixing entropy.

"Double percolation" formed by co-continuous structure of immiscible polymer has been investigated to achieve higher electrical conductivity with reduced conductive filler content (carbon black[6-8], carbon nanotubes[9-13], graphene[14-17], *etc.*) and decreased the percolation threshold[18]. Some researchers have also found that double percolation and the morphology of conductive fillers in immiscible polymer blend had a significant influence on dielectric properties of nanocomposites[19]. For example, Yuan *et al.*[20] reported that MWCNTs/LDPE/PVDF composites displayed significantly reduced percolation threshold (5.7 vol%), high permittivity (~ 500) which can be attributed to a double percolated structure. Meanwhile, it has been reported that the different factors during the fabrication process (*e.g.*, mixing approach[21], shearing forces[22] and mixing time[23]) to material-related (*e.g.*, fillers size, viscosity ratio, and intrinsic properties)[24] can bring forth perturbation of the thermodynamic distribution of fillers at equilibrium and cause the migration of fillers which also may has a remarkable influence on the dielectric properties of the composite[25]. Xu *et al.* [26] reported that when MWCNTs were completely accommodated inside the PA6 after migration from the PPS, the resistivity of the final blend nanocomposite increased relative to when MWCNTs were at the interface between PPS and PA6. However, this line of research to date lacks a refined control of the overall migration degree, and the influence of different migration degrees on the microwave dielectric properties remain unclear. On the other hand, it has been reported that as mixing of an immiscible pair of polymers can produce an assortment of stage morphologies depending on concentration of constituent parts and the qualities of the segments, blend morphology may vary from irregular scattering of nearly sea-island to continuous morphology [27, 28]. It provides a novel idea to tune the dielectric properties by controlling the morphology of the matrix. Consequently, the immiscible polymer-based nanocomposites afford an ideal platform to tune the microwave dielectric properties adequately.

In this paper, we chose CNTs as the nanofillers which can facilitate the migration process and the formation of percolation. We exploited the driving force of the system toward thermodynamic equilibrium state to successfully control the migration degree at the interface of CNTs from the perspective of dynamics, and analyzed in detail the influence and mechanism of migrating scenarios on the microwave dielectric properties as compared with the aforementioned existing work. In addition to the regulation of the filler phase, we further explored the role of the immiscible matrix. Due to the varying size of phase domain spontaneously formed by the immiscible polymer system, the multiple microwave dielectric relaxation processes can be constructed. We have provided unique insights on the relationship between the position of the relaxation peak and the phase domain distribution. This paves a new way for the design of broadband microwave absorbing materials from the perspective of nanocomposite matrix, which is different from most researches constructing multiple absorbing effects through the complex hierarchical design of adding phases. Also, by combining the CNTs interface localization control with the matrix co-continuous structure design, we found that the interface promotes double percolation effect to achieve percolation at significantly low CNTs loading, which utilized immiscible structural advantages and facile dynamic control methods. The rest of paper is organized as follows.

## 2. Experimental

*2.1 Materials*

Polystyrene (PS) resin pellets with General-purpose type I and Polymethyl methacrylate (PMMA) pellets with high flow injection stage were received from Aladdin Shanghai (China). The grade of PMMA and PS is extrusion grade. The MFR of PMMA and PS in 200°C/5kg is 4.8 and 7.2 g/min. Pristine CNTs (purity: 95%, outer diameter: < 8 nm, inner diameter: 2-5 nm, length: 10-30 μm) were obtained from Chengdu Organic Chemicals Co.Ltd., Chinese Academy of Sciences. The average size of CNT agglomerates is ~20 μm.

*2.2 Nanocomposites preparation*

The polymers were dried at 60 °C for 24 h to remove moisture before use. The (PMMA-CNTs)/PS and PMMA/PS-CNTs nanocomposites were prepared in a two-step process. Firstly, CNTs and PMMA or CNTs and PS were mixed on Miniature High-performance Composite Material Mixing Molding System (HAAKE MiniLab II) at 200 °C and 60 rpm for 5 min to obtain PMMA-CNTs and PS-CNTs masterbatches. Secondly, the PMMA-CNTs masterbatch was blended with PS at 200 °C and 60 rpm for 2.5, 5.0 and 7.5 min to obtain nanocomposites with different CNTs migration degrees named as (PMMA-CNTs)/PS-1, (PMMA-CNTs)/PS-2 and (PMMA-CNTs)/PS-3 respectively. The PS-CNTs masterbatch was blended with PMMA at 200 °C and 60 rpm for 5 min to obtain PMMA/(PS-CNTs) nanocomposites. In this work, the volume ratios of PMMA to PS were set at 50/50, 80/20 and 20/80 to obtain blends with two kinds of morphologies, *i.e.*, co-continuous morphology and sea-island morphology.

*2.3 Morphology characterization*

The cold field-emission scanning electron microscopy (FE-SEM, Hitachi S-4800) with accelerating voltage of 3 kV was employed to analyze the morphology of the samples. The samples were first fractured with liquid nitrogen treatment and then a selective solvent dissolution technique was used to clearly distinguish the two-phase[29]. Samples of 0.2 g were immersed in a large volume of formic acid and stirred gently at 40 °C to selectively dissolve PMMA component and ultrasonically cleaned with ethanol. Subsequently, the dried samples were sputter-coated with a thin layer of gold before SEM observation. TEM measurements were performed on a HT-7700 transmission electron microscopy operated at an acceleration voltage of 100 kV to analyze the localization and migration of the CNTs. The specimens were cut into films with a thickness of 50 nm by Ultra-thin frozen slicer (EMUC7) equipped with a glass knife.

*2.4 Dielectric measurements*

Dielectric measurements in the frequency range of 8.2-12.4 GHz were performed by vector network analyzer (R&S, ZNB20) and the complex permittivities were extracted from the S-parameters by Nicolson-Ross-Weir method, which is widely used as derived from Maxwell's equations, constitutive equations and boundary conditions[30]. The cuboid shaped specimens (22.86 × 10.16 × 2 mm) compressed at 200 °C were used.

*2.5 Electrical conductivity measurements*

Compression molded samples (20 × 20 × 1 mm) at 200°C were prepared for electrical conductivity measurements. DC electrical conductivities at room temperature were recorded using a CHI 660E electrochemical analyzer from CH Instruments by linear sweep voltammetry (LSV).

*2.6 Contact Angle Measurements*

Contact angles of neat polymers (PMMA and PS) were evaluated in a sessile drop mold with Video-based contact angle measuring device (OCA20, Germany). The neat polymer pellets were compression molded between clean silicon wafers at 200 °C. The molded samples were square-shaped with a diameter of 100 mm and 2 mm thickness. Contact angles were measured on 3.5 μL of wetting solvent (propanediol and distilled water) at room temperature.

After introducing the experimental details, we first studied the effect of CNTs migration and localization on the microwave dielectric properties at Sec.3.1 and 3.2, Then the effect of immiscible blends morphologies phases on dielectric properties are detailed in Sec.3.3. Sec 3.4 is devoted to the interface-promoted double percolation phenomenon. The major outcomes are summarized in Conclusions.

## 3. Results and discussions

*3.1 Localization in the Equilibrium State and Migration of CNTs*

From the results of shear viscosity, it can be known that under high shear force, the viscosity ratio of PMMA and PS is close to 1 (Fig. S1). Herein, the localization of CNTs in the polymer blends is governed by thermodynamic equilibrium and it can be predicted by calculating the wetting coefficient $\omega$ based on Young's equation (Eq (1)), proposed by Sumita *et al*.[8] as follows:

$$\omega = \frac{\gamma_{CNTs-PS} - \gamma_{CNTs-PMMA}}{\gamma_{PMMA-PS}} \tag{1}$$

Where $\gamma_{CNTs-PMMA}$, $\gamma_{CNTs-PS}$ and $\gamma_{PMMA-PS}$ are the interfacial energies between CNTs and PMMA, CNTs and PS, and PMMA and PS, respectively. If $\omega > 1$, CNTs will be preferentially distributed in PMMA phase; if $\omega < -1$, CNTs will be expected to localize in PS phase; and if $-1 < \omega < 1$, then CNTs will be selectively localized at the interface between PMMA and PS phase.

Here we use geometric equation (Eq (2a))[31] and harmonic equation(Eq(2b))[32] to determine the $\gamma_{CNTs-PMMA}$, $\gamma_{CNTs-PS}$ and $\gamma_{PMMA-PS}$ as the geometric mean method is more accurate for nonpolar/polar systems and harmonic equation for nonpolar/nonpolar systems.

$$\gamma_{12} = \gamma_1 + \gamma_2 - 2(\sqrt{\gamma_1^D \gamma_2^D} + \sqrt{\gamma_1^P \gamma_2^P}) \tag{2a}$$

$$\gamma_{12} = \gamma_1 + \gamma_2 - 4(\frac{\gamma_1^D \gamma_2^D}{\gamma_1^D + \gamma_2^D} + \frac{\gamma_1^P \gamma_2^P}{\gamma_1^P + \gamma_2^P}) \tag{2b}$$

Where $\gamma_{12}$ is the interfacial energy of component 1 and 2, and $\gamma^D$, $\gamma^P$ represent the dispersive and polar contributions, respectively.

The surface energy, dispersion, and polar contributions of the neat polymers are estimated from the contact angle data using Fowkes' method (Eq (3)) [33]

$$\cos\theta = 2\frac{\sqrt{\gamma_L^D}\cdot\sqrt{\gamma_S^D}}{\gamma_L} + 2\frac{\sqrt{\gamma_L^P}\cdot\sqrt{\gamma_S^P}}{\gamma_L} - 1 \tag{3}$$

where $\theta$ is the contact angle; $\gamma_L$ and $\gamma_S$ represent the liquid and solid surface tension, respectively.

Table 1 lists the surface free energies calculated for PS, PMMA and CNTs at 200 °C as per Guggenheim equations[34]. The interfacial tensions between the different blend components can be calculated based on Eq. (2), and the results are shown in Table 2. From Eq (1), wetting coefficient is estimated to be -1.42, which suggests that CNTs prefer to dwell in the PS phase of the blends[8].

**Table 1.** Surface Free Energies of the Polymers and CNTs at 200 °C

| Material | surface free energy(mN/m) | | |
|---|---|---|---|
| | $\gamma$ | $\gamma^D$ | $\gamma^P$ |
| PMMA | 38.1 | 34.4 | 3.7 |
| PS | 31.5 | 29.7 | 1.8 |
| CNTs[35] | 27.8 | 17.6 | 10.2 |

**Table 2.** Interfacial energies and wetting coefficient at 200 °C

| Components | Interfacial tension(mN/m) | | | $\omega$ |
|---|---|---|---|---|
| | $\gamma_{PMMA-PS}$ | $\gamma_{CNTs-PMMA}$ | $\gamma_{CNTs-PS}$ | |
| Geometric equation | 1.29 | 8.48 | 6.65 | -1.42 |

It can be found in Fig. 1 that CNTs are migrating to the PS phase which has better affinity when PMMA-CNTs masterbatch blends with PS phase[18, 36]. It can be seen from Fig. 1(a) that the CNTs have disappeared from the PMMA phase of the masterbatch, and most of them distribute in the interface and the PS phase (the black part in PMMA phase is not agglomerates of CNTs, but dispersed PS phase). It can be found after zooming in the interface that some CNTs diffuse towards the PS phase and their directions are perpendicular to the interface (Fig. 1(b)). The works of Göldel *et al*. [37] suggested that the transfer dynamics as well as the stability of different nanofillers at the interface reveal a strong dependence on the fillers' aspect ratio. Low interfacial stabilities and high transfer speeds between the blend phases can be deduced for fillers with very high aspect ratios, entitled as the "Slim-Fast Mechanism" (SFM). The curvature of the interface induced by the CNTs (rod-like) is independent of their movement toward the PS phase. The driving force does not wane during the migration until CNTs completely migrated to the PS phase and the transfer should thus be faster and more effective. Meanwhile, perpendicular direction can maintain maximum aspect ratio and driving force. This provides a reasonable explanation for the migration of CNTs from the PMMA phase into PS phase. Fig. 1(c) displays that CNTs migrated to the PS phase have the advantage of low agglomeration, it follows that the migration process assists in yielding blends with better dispersion of CNTs in the PS phase.

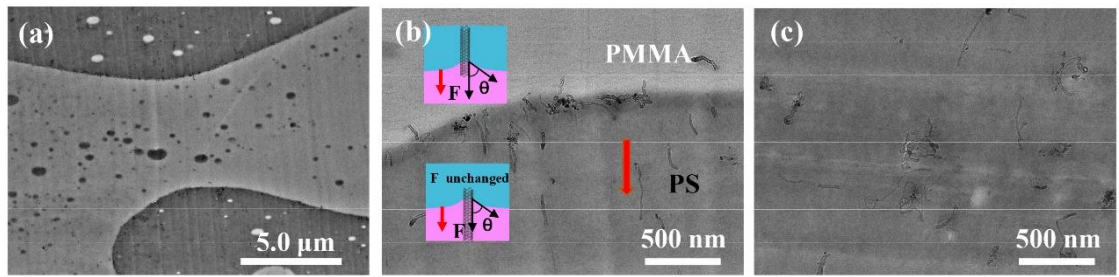

**Fig. 1**. TEM images of the (PMMA-CNTs)/PS (50/50) with 0.53 vol% CNTs. The dark and light parts in the TEM images correspond to the PS and PMMA phases, respectively.

*3.2 Selective localization of CNTs in blends: Effect on dielectric properties*

The thermodynamic equilibrium can be disturbed by processing conditions and dynamics factors during the melt mixing stage of polymer compounding which provides an ingenious approach to nanocomposites design. We realize a fine control of selective localization of CNTs by regulating the migration through the dynamics perspective and the masterbatch technique as shown in Fig. 2(a). A typical co-continuous structure can be observed in Fig. 2(b) of which volume ratio of PS and PMMA phase is controlled at 50/50. For more morphological characterization, see Fig.

S2. CNTs will migrate from PMMA to PS phase when PMMA-CNTs is used as the masterbatch as described in section 3.1, and the migration can be regulated through processing time. Herein, we managed to realize several typical CNTs migrating scenarios. Fig. 2(d) shows that most of the CNTs are preferentially located at the interface of PMMA and PS phases and only a small amount are dispersed in the PS phase when the mixing time is 2.5 min ((PMMA-CNTs)/PS-1) (mixing time refers to the time after adding PS into PMMA-CNTs masterbatch). This is because most of the CNTs have not sufficient migration time to disperse. It can be seen in Fig. 2(e) that the migration of CNTs becomes stronger as fewer CNTs are found at the interface, and more are in the PS phase with the melt mixing time extended to 5.0 min ((PMMA-CNTs)/PS-2). Fig. 2(f) presents the morphology with further CNT migration after 7.5 min ((PMMA-CNTs)/PS-3) of mixing and CNTs almost disperse in the PS phase, which is in good agreement with the above theoretical calculation result predicting the thermodynamics dominated localization of CNTs. However, when PS-CNTs is used as the masterbatch, the migration of CNTs will not occur. Meanwhile, it can be found in Fig. 2(g) that the aggregation of CNTs in PMMA/(PS-CNTs) (50/50) is more serious due to the lack of CNTs migration process. By selectively extracting the PMMA phase, we further verify the migration degree and distribution of CNTs from a macroscopic perspective. As shown in Fig. 2(c), as the CNTs migrate to the PS phase, the number of CNTs at the interface and PMMA phase decrease, thus, the color of the extraction solutions becomes lighter.

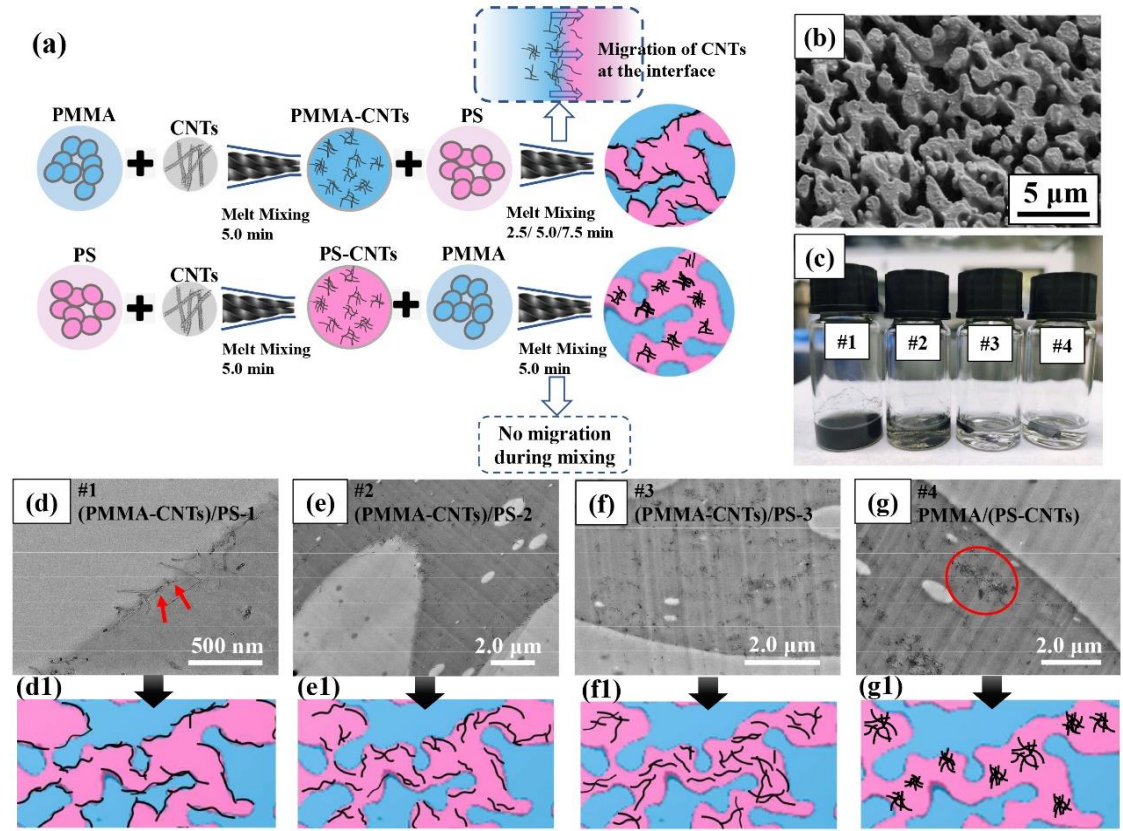

**Fig. 2**. (a) Schematic representations showing the preparation process and migration evolution of the CNTs; (b) SEM images of (PMMA-CNTs)/PS-2 filled with 0.53 vol% CNTs where the PMMA phase was etched by formic acid; (c) Solvent extraction solutions of nanocomposites after PMMA phase extraction; TEM images of nanocomposites filled with 0.53 vol% CNTs: (d) (PMMA-CNTs)/PS-1 (the red arrows mark the CNTs located at the interface); (e) (PMMA-CNTs)/PS-2; (f) (PMMA-CNTs)/PS-3; (g) (PMMA-CNTs)/PS (the red circle marks the agglomerates of CNTs in the PS phase); (d1) ~ (g1) schematic diagram for the location and dispersion of CNTs in four kinds of nanocomposite respectively.

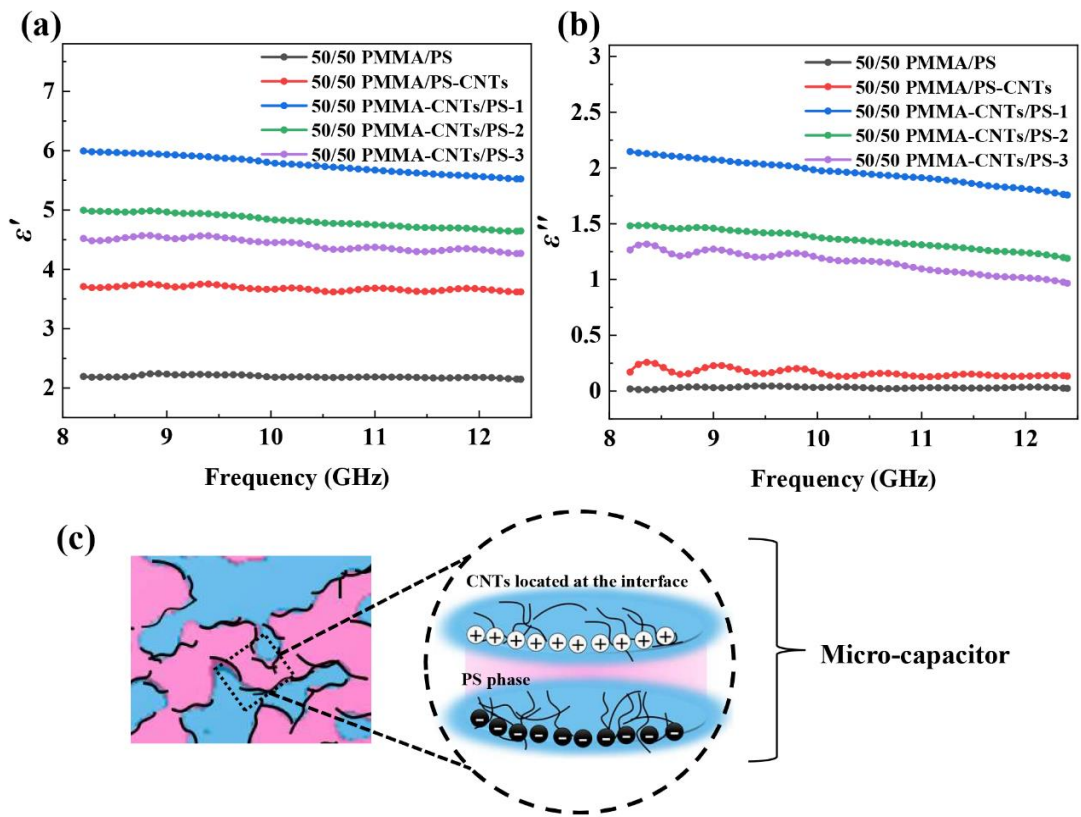

**Fig. 3**. Variation of complex dielectric frequency spectra (8.2-12.4 GHz) for (PMMA-CNTs)/PS-x filled with 0.53vol% of CNTs: (a) real part; (b) imaginary part; (c) schematic diagram of micro-capacitor.

The complex dielectric frequency spectra of the nanocomposites are observed in the frequency range of 8.2-12.4 GHz at room temperature. Fig. 3(a) displays the frequency dispersion of the real part of the permittivity ($\varepsilon'$). One can see that (PMMA-CNTs)/PS-1 (50/50) with the highest concentration of CNTs at the interface along with the shortest mixing time has the maximum $\varepsilon'$ value. This finding can be explained by the existence of micro-capacitor networks which offers the essential physical mechanism for the critical behavior of the permittivity near percolation[38]. The micro-capacitors[39] are formed by the neighboring CNTs networks at the interface as electrodes and a very thin layer of matrix in between as dielectric as illustrated in Fig. 3(c). Each micro-capacitor contributes an abnormally large capacitance, which can be correlated with the significant increase in the value of $\varepsilon'$ [40]. With the extension of processing time, the concentration of CNTs at the interface tends to be lower and the formation of migration reduces the number of micro-capacitors formed by the neighboring CNTs networks at the interface. Normally, the ability to store charges depends on the area of the plate (S), the distance between the plates (d) and the charge of the capacitor (Q). The CNTs networks connected at the interface (partially connected, not percolated) have larger Q and S than the single CNT in the PS phase. As a result, the value of $\varepsilon'$ decreases with the processing time. Fig. 3(b) shows frequency dispersion of the imaginary part of permittivity ($\varepsilon''$) which represent the dielectric loss. The nanocomposite displays an increase of conductivity when the conductive fillers selectively build up their conductive pathway at the interface of a co-continuous polymer blend, as will be elucidated in detail in Sec 3.4. As such, the localization of CNTs at the interface leading to conductive loss in the alternating electric field is the main reason for the high dielectric loss. It can be found that (PMMA-CNTs)/PS-1 (50/50) which builds a more complete conductive network has the largest $\varepsilon''$. Due to the further migration of CNTs, the content of CNTs in the PS phase increases and conductive pathway at the interface is gradually destroyed so $\varepsilon''$ of (PMMA-CNTs)/PS-2 (50/50) and (PMMA-CNTs)/PS-3 (50/50) reduce accordingly. For PMMA/(PS-CNTs) (50/50), due to the agglomeration the number of dipole polarization centers formed by CNTs clusters reduces and the effect of CNTs on improving the $\varepsilon'$ value is also weakened[41]. Compared with (PMMA-CNTs)/PS (50/50), PMMA/(PS-CNTs) (50/50) can only form local conductive networks in the CNTs clusters and hence results in significant drop of dielectric loss ($\varepsilon''$), as shown in Fig. 3(b). For comparison, the complex dielectric frequency spectra of PMMA-CNTs and PS-CNTs composites is shown in Fig. S3. In addition, we also studied the different location of CNTs in the droplet morphology nanocomposite. The microstructure characterization and dielectric spectrum are shown in Fig. S4. The above results suggest that via the location and dispersion control of CNTs, one can modulate the microwave dielectric performance as desired, which is very useful for the practical applications that usually pose specific requirement on the dielectric constant and loss.

*3.3 Effect of immiscible blends morphologies phases on dielectric properties*

As important as the CNTs migrating scenarios, the unique features of immiscible polymer matrix also play important role in formulating the microwave dielectric responses. One function of the PMMA/PS matrix has been demonstrated to aid the CNTs migration as discussed above. Another key function of this matrix is to formulate some specific phase domain structures, which will have significant influence on the microwave dielectric properties of the nanocomposites as will be detailed below. To reveal the influence of introducing unique morphologies and blend interface into nanocomposites on dielectric response, we built two kinds of morphologies systems with varied volume ratio of PMMA and PS phase. In this part of work, we fabricated PMMA/(PS-CNTs) in which CNTs selectively locate in the PS phase as shown in Fig. 4(a). SEM and TEM images show that we have successfully obtained the co-continuous morphology by controlling the volume ratio of PMMA/PS to 50/50 (Fig. 4(c-c1)) and the sea-island morphology by controlling the volume ratio to 80/20 and 20/80 (Fig. 4(b-b1, d-d1)).

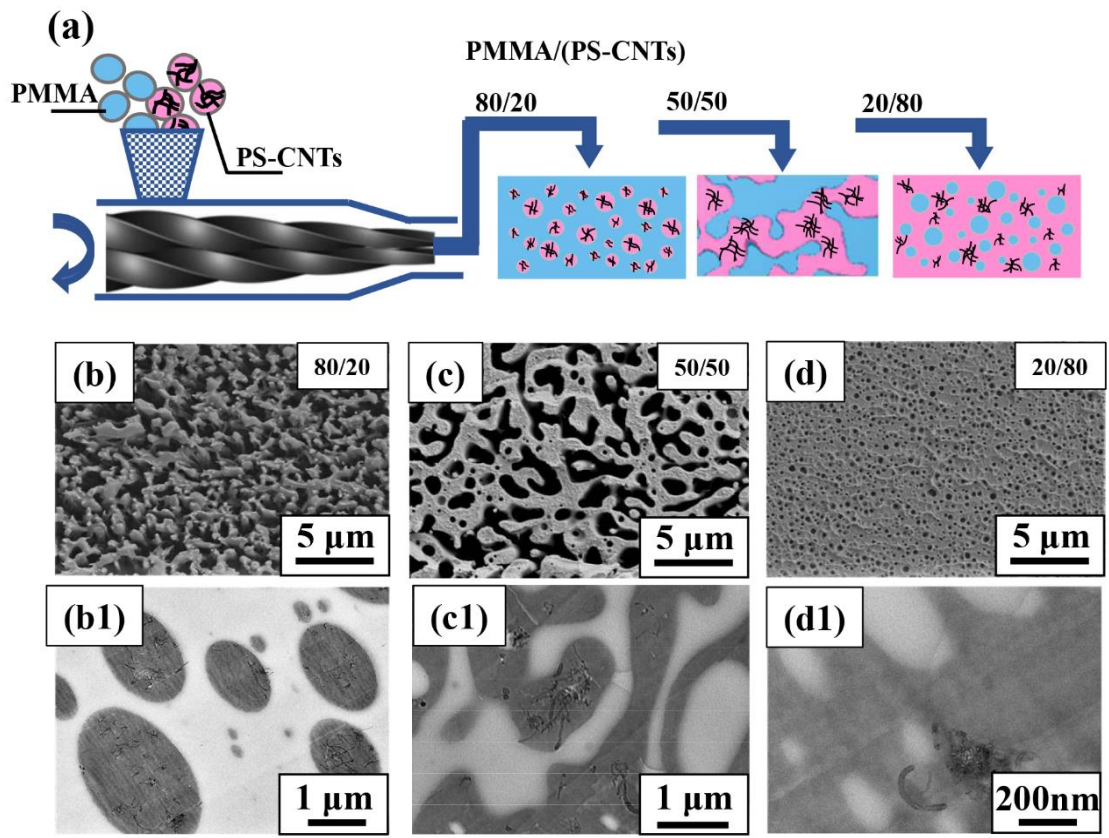

**Fig. 4**. (a) Schematic representations showing the composition mixing ratio and the localization of the CNTs in different morphologies of nanocomposites; SEM images of the PMMA/(PS-CNTs) filled with

0.53vol% of CNTs: (b) PMMA/(PS-CNTs) (80/20), (c) PMMA/(PS-CNTs) (50/50) and (d) PMMA/(PS-CNTs) (20/80), where the PMMA phase was etched by formic acid; TEM images of (b1) PMMA/(PS-CNTs) (80/20), (c1) PMMA/(PS-CNTs) (50/50), (d1) PMMA/(PS-CNTs) (20/80).

Fig. 5(a) presents the complex dielectric frequency spectra of the PMMA/(PS-CNTs) with the above two morphologies in the frequency range of 8.2-12.4 GHz at room temperature. In order to visualize the difference in phase domain size, statistical analyses on the average area of domain size were carried out with the *ImageJ* software. Fig. S5(a1-c2) shows the phase domain sizes statistics. Both of two morphologies have phase domain structures of varying size distribution. We added up the perimeters of the interface in the two-dimensional images to qualitatively count the interface areas of the nanocomposites (Table. S1). The statistics of interface areas approximately indicate that the sea-island morphology (80/20; 20/80) produces more phase interface than co-continuous morphology (50/50). With the ratio of PS phase decrease, the concentration of CNTs in PS phase increase correspondingly based on the excluded volume effect. As a result, the differences in the conductivity and permittivity between the two phases are enlarged. The interfaces of the PMMA and PS phase domain limit the movement of charges which results in the restriction and accumulation of bound charges at the interface and causes the interfacial polarization. The addition of this kind of new polarization mechanism can improve the $\varepsilon'$ value so PMMA/(PS-CNTs) (80/20) which has the largest interface area and the biggest difference in polarization ability between two phases has the highest $\varepsilon'$ value (Fig. 5(a)). For the imaginary part of permittivity, we can see the PMMA/(PS-CNTs) (80/20) has the maximum value (Fig. 5(b)). The dielectric loss is mainly due to interfacial polarization relaxation loss and leakage conduction loss of local conductive network formed by CNTs clusters: PMMA/(PS-CNTs) (80/20) generate more interfacial polarization at interfaces promoting a larger interfacial polarization loss, and the CNTs trapped in the PS phase form large number of conductive networks which can cause leakage conduction loss. For PMMA/(PS-CNTs) (50/50) and PMMA/(PS-CNTs) (20/80), these two types of loss effect have different degrees of weakening.

Another observation is the emergence of relaxation peaks at different frequencies in the dielectric spectrum. For PMMA/(PS-CNTs) with selective localization of CNTs in PS phase, distinguishable Cole-Cole semicircles are observed (Fig. 5(c1-c3)), suggesting that there occurred multiple dielectric relaxation processes, which can be ascribed to the phase domain structures of size distribution in the immiscible polymer blends[42]. In general, under the alternating electric field, the relaxation processes in a dielectric medium are caused by the delay of induced charges. In the immiscible polymer system, large amount of PS and PMMA phase interfaces have been introduced which can cause significant interfacial polarization. Meanwhile, owing to the selective localization of CNTs in PS phase, the difference in polarization abilities between the two phases becomes larger, so the dielectric relaxation processes caused by two-phase interfaces become more obvious[5]; this explains why the relaxation peaks are becoming more apparent with the migration of CNTs to PS phase in Fig. 3(b). As illustrated in Fig. 5(b), we notice that three kinds of nanocomposite with different volume ratios of PMMA and PS phase have different relaxation response position and intensity. Meanwhile, it is observed that the relaxation peaks of PMMA/(PS-CNTs) (20/80) and PMMA/(PS-CNTs) (80/20) having the same sea-island morphology are located in a similar frequency band. In comparison, relaxation peaks of PMMA/(PS-CNTs) (50/50) with co-continuous morphology exhibits an obvious red shift and enhancement as a result of the larger dipole size[43] contributed by the increase of the phase domain size (see Fig. S5). Therefore, it is reasonable to conclude that the difference in dielectric behavior is related to their difference in morphology and structural feature, and that the dielectric relaxation of different frequency bands can be realized by controlling the phase domain size and distribution. Based on the Ostwald ripening and Coarsening mechanism[44], it is feasible to further regulate relaxation by changing the domain size and distribution through heat treatment[45]. This paves a new way for the design of broadband microwave absorbing materials from the perspective of nanocomposite polymer matrix instead of absorbing fillers.

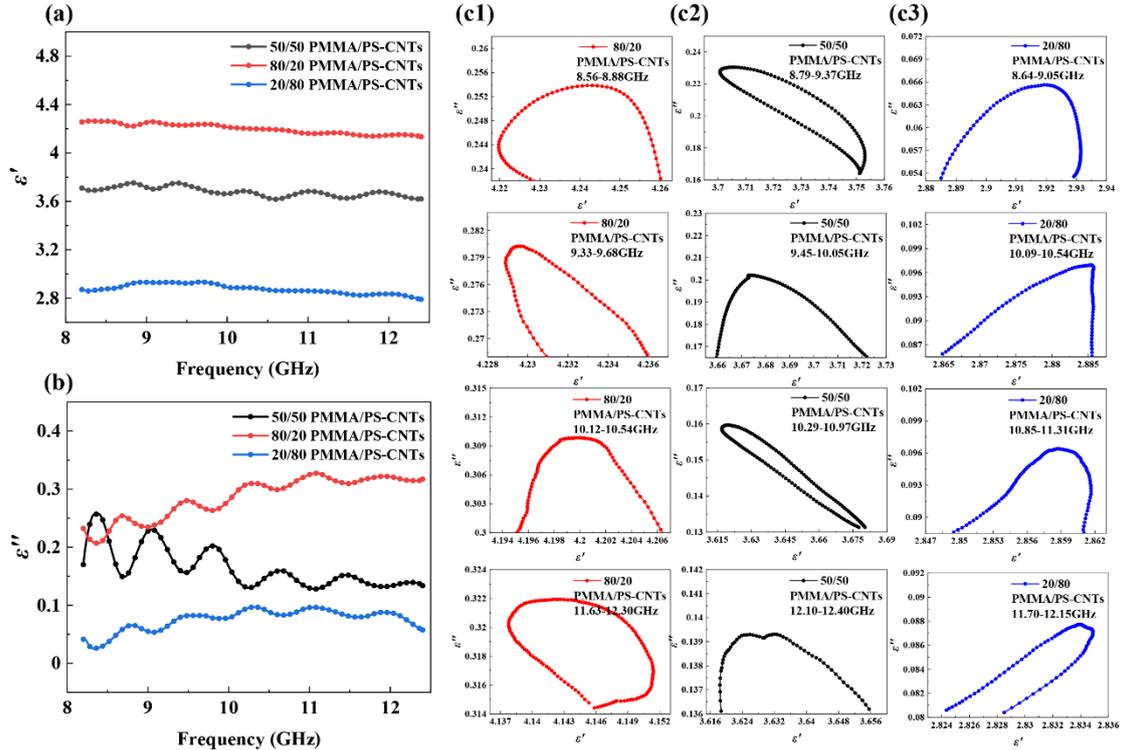

**Fig. 5.** Variation of complex dielectric frequency spectra (8.2-12.4 GHz) for PMMA/(PS-CNTs) with different volume ratio of PMMA and PS: (a) real part; (b) imaginary part; typical Cole-Cole semicircles of (c1) PMMA/(PS-CNTs) (80/20), (c2) PMMA/(PS-CNTs) (50/50) and (c3) PMMA/(PS-CNTs) (20/80) in different frequency bands.

*3.4 Effect of content of CNTs in blends on conductive percolation*

From section 3.2 and 3.3, it is found that CNTs are more enriched at the interface with the mixing time of 2.5 min and a lower interface area along with the co-continuous structure can be obtained with the matrix ratio of 50/50. This scenario affords a possible way to further enhance the so-called double percolation effect by interface to achieve percolation at low CNTs loading, as will be elucidated below. Fig. 6(a, b) shows the variation of complex dielectric frequency spectra in the frequency range of 8.2-12.4 GHz at room temperature for unfilled and different content of CNTs filled (PMMA-CNTs)/PS-1 nanocomposites. It is seen $\varepsilon'$ rises with increase of concentration of CNTs and it decreases with increasing frequency. Especially, there is a significant increase in permittivity value for 1.06 vol% of CNTs loading. At low concentration of CNTs loading, the CNTs are isolated and there is no interaction between them. When the CNTs content is increased the CNTs tend to approach each other and form an interconnected cluster. Under this condition, the polarization of a single nanotube can have influence on orientation of adjacent dipoles, which leads to greater average polarization and increased $\varepsilon'$ [46-47]. When the content of CNTs reaches 1.06 vol%, based on the co-continuous structure, the CNTs conductive network at the interface becomes more complete which form double percolation structure. If the concentration approaches the percolation threshold the $\varepsilon'$ will have a sharp increase, which is result of the significant increase in $\varepsilon'$ for 1.06 vol% of CNTs loading (Fig. 6(c)). On the other hand, it can be found in Fig. 6(b) that with the increase of CNTs loading, the imaginary part of permittivity ($\varepsilon''$) rises correspondingly and there is also a sharp increase in $\varepsilon''$ for 1.06 vol% of CNTs loading. The dielectric loss consists of two parts: one is due to polarization relaxation, and the other originates from the leakage current loss. As the content of CNTs increases, the conductive networks formed by CNTs at the interface become more complete, which greatly increases the leakage conduction loss.

Fig. 6(d) shows the DC conductivity values of (PMMA-CNTs)/PS-1 (50/50) with different content of CNTs filled. When the content of CNTs is low, with the increasing content of CNTs, the conductivity values increase slightly further. Percolation structure has not been achieved with lower filler loadings.

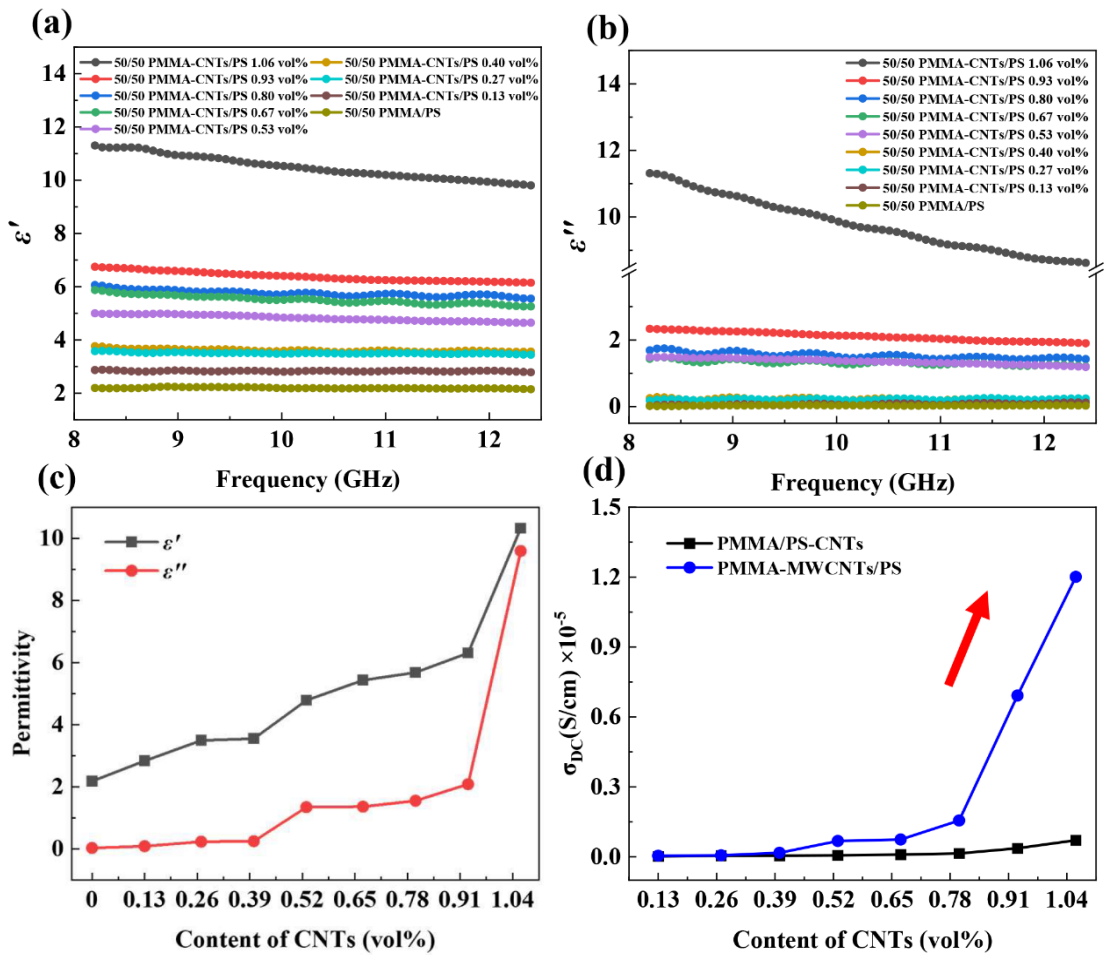

**Fig. 6**. Variation of complex dielectric frequency spectra (8.2-12.4 GHz) for (PMMA-CNTs)/PS-1 (50/50) with different concentration of CNTs: (a) real part; (b) imaginary part; (c) Real parts and imaginary parts of permittivity of (PMMA-CNTs)/PS (50/50) with different CNTs loadings at 10.5 GHz; (d) DC conductivity of the (PMMA-CNTs)/PS (50/50) and PMMA/(PS-CNTs) (50/50) with different concentration of CNTs.

However, when the concentration of CNTs reaches 1.06 vol% and forms more complete conducting pathways in the interface, the conductivity increases dramatically. The connection of the interface phase and the percolation of the CNTs form a double percolation structure. By contrast, with the same loading of CNTs, (PMMA-CNTs)/PS (50/50) have larger conductivity values than PMMA/(PS-CNTs) (50/50). Even as the CNTs concentration reaches 1.06 vol%, PMMA/(PS-CNTs) (50/50) still do not show a significant increase in conductivity values. The TEM images of (PMMA-CNTs)/PS (50/50) and PMMA/(PS-CNTs) (50/50) with 1.06 vol% CNTs loading are shown in Fig. S6. It can be concluded that (PMMA-CNTs)/PS (50/50) with CNTs at the interface achieves the best conductivity with low percolation threshold by forming the double percolated structure throughout the blend. In other words, the interface promoting percolation is the most effective double percolation structure. Indeed, some researchers also managed to control nanofillers distributed at the interface to lower the percolation. Mohammed H. *et al*.[48] introduced styrene–butadiene–styrene (SBS) tri-block copolymer in the PP/PS/CB blend which was selectively localized at the interface between PP and PS phase. There is a 40% reduction in the percolation threshold in the (70/30) PP/PS blend upon addition of 5 vol% SBS because of the higher affinity of CB to the polybutadiene (PBD) section of the SBS copolymer. Eyal *et al*.[49] introduced pyridine-modified poly(ethylene-co-methacrylic acid) as the minor component that can form strong interactions with the CNTs via π-π interactions and confined the percolated network at the polyamide/polypropylene interface. This ternary structure has lower electrical resistivity as compared to co-continuous binary composites. However, their work requires the introduction of the third phase and complex chemical treatment. In comparison, the present work has the absolute advantage of facile processing. Note that the present threshold percolation has not reached a satisfying level, which can be attributed to the following reasons: a) the main reason is the limitations of processing equipment which cannot ensure CNTs all located at the interface instead of migrating to the PS phase, and hence the unwanted CNTs in the PS phase increase the percolation; b) another possible reason could be the melt mixing processing causes damage to the CNTs and reduces the aspect ratio[50]. To address these two issues, we could consider an updated facility with more precise control of processing time and use CNTs with large aspect ratio and the relevant work is underway. It should also be noted that, more practical processing method may demand longer processing time, exploitation of which shall be considered in the future work. Even so, this work fully demonstrates the effectiveness of the interface promoting percolation effect with such a facile preparation technique.

## 4. Conclusions

To summarize, we used the driving force of an immiscible polymer nanocomposite system toward the thermodynamic equilibrium state to control the migration degree of CNTs via the interface, revealing the influence of CNTs localization on microwave dielectric properties. The TEM images present distinctly several migrating scenarios of CNTs from the PMMA phase to PS phase through the control of processing. With the migration of CNTs to PS phase, the complex permittivity decreases correspondingly owing to the suppression of micro-capacitors structure. Furthermore, two kinds of morphology (co-continuous and sea-island) were obtained by adjusting the ratio of two phases. We found that the multiple dielectric relaxation processes are produced by selective localization of CNTs and phase domain structures of varying size in two different morphologies nanocomposites, which has great value for the design of broadband microwave absorbing materials. We explored in-depth the influence of co-continuous structure and CNTs distribution on dielectric properties and electric conductivity. When the CNTs are located at the two-phase interface, the greatest increase in electrical conductivity and dielectric constant is found for (PMMA-CNTs)/PS (50/50) filled with 1.06 vol% of CNTs, which is attributed to the interface promoting double percolation effect. Compared with the sheer polymer and multi-fillers nanocomposite, the immiscible polymer system provides a novel composite design method with more structural adjustability and tunable dielectric properties, which further enrich the "plainified" design methodology of nanocomposite materials for a spectrum of microwave engineering applications such as microwave absorbers.

**Acknowledgements**

This work is supported by ZJNSF No. LR20E010001 and National Key Research and Development Program of China No. SQ2018YFE011526 and Zhejiang Provincial Key Research and Development Program (2019C01121).